\documentclass[a4paper,UKenglish]{lipics-v2018}
\usepackage{microtype}%if unwanted, comment out or use option "draft"

\newcommand\nc{\newcommand}
\nc{\pf}[1]{ \noindent \emph{Proof.} #1
 \hfill \qed\par}

\title{Bounded-Degree Cut is Fixed-Parameter Tractable}

\titlerunning{Bounded-Degree Cut is FPT}%optional, please use if title is longer than one line

\author{Mingyu Xiao}{School of Computer Science and Engineering,
University of Electronic Science and Technology of China, Chengdu, China}{myxiao@gmail.com}{https://orcid.org/0000-0002-1012-2373}{Supported by the National Natural Science Foundation of China, under
grants 61772115 and 61370071.}%mandatory, please use full name; only 1 author per \author macro; first two parameters are mandatory, other parameters can be empty.

\author{Hiroshi Nagamochi}{Department of Applied Mathematics and Physics,
  Graduate School of Informatics, Kyoto University, Japan}{nag@amp.i.kyoto-u.ac.jp}{}{}

\authorrunning{M.\,Xiao and H.\,Nagamochi}%mandatory. First: Use abbreviated first/middle names. Second (only in severe cases): Use first author plus 'et al.'

\Copyright{Mingyu Xiao and Hiroshi Nagamochi}%mandatory, please use full first names. LIPIcs license is "CC-BY";  http://creativecommons.org/licenses/by/3.0/

\subjclass{G.2.2 Graph Theory}% mandatory: Please choose ACM 2012 classifications from https://www.acm.org/publications/class-2012 or https://dl.acm.org/ccs/ccs_flat.cfm . E.g., cite as "General and reference $\rightarrow$ General literature" or \ccsdesc[100]{General and reference~General literature}.

\keywords{FPT; Important Cuts; Graph Cuts; Graph Algorithms}%mandatory

\category{}%optional, e.g. invited paper

\relatedversion{}%optional, e.g. full version hosted on arXiv, HAL, or other respository/website

\supplement{}%optional, e.g. related research data, source code, ... hosted on a repository like zenodo, figshare, GitHub, ...

\funding{}%optional, to capture a funding statement, which applies to all authors. Please enter author specific funding statements as fifth argument of the \author macro.

\acknowledgements{%I want to thank \dots
}%optional

\EventEditors{Ioannis Chatzigiannakis, Christos Kaklamanis, D\'{a}niel Marx, and Don Sannella}
\EventNoEds{4}
\EventLongTitle{45th International Colloquium on Automata, Languages, and Programming (ICALP 2018)}
\EventShortTitle{ICALP 2018}
\EventAcronym{ICALP}
\EventYear{2018}
\EventDate{July 9--13, 2018}
\EventLocation{Prague, Czech Republic}
\begin{document}

\maketitle

\begin{abstract}
In the bounded-degree cut problem, we are given a  multigraph $G=(V,E)$, two disjoint vertex subsets $A,B\subseteq V$, two functions  $\mathrm{u}_A, \mathrm{u}_B:V\to \{0,1,\ldots,|E|\}$ on $V$, and an integer $k\geq 0$. The task is to determine whether there is a minimal $(A,B)$-cut $(V_A,V_B)$ of size at most $k$ such that the degree of each vertex $v\in V_A$ in the induced subgraph $G[V_A]$ is at most $\mathrm{u}_A(v)$ and the degree of each vertex $v\in V_B$ in the induced subgraph $G[V_B]$ is at most $\mathrm{u}_B(v)$. In this paper, we show that the bounded-degree cut problem is fixed-parameter tractable by giving a $2^{18k}|G|^{O(1)}$-time algorithm. This is the first single exponential FPT algorithm for this problem. The core of the algorithm lies two new lemmas based on important cuts, which give some upper bounds on the number of candidates for vertex subsets in one part of a minimal cut satisfying some properties. These lemmas can be used to design fixed-parameter tractable algorithms for more related problems.
 \end{abstract}

\section{Introduction}\label{sec:intro}
A \emph{cut} of a graph is a partition of the vertices of the graph into two disjoint subsets.
Graph cuts play an important role in combinatorial optimization and graph theory.
The classical \emph{minimum cut problem} is well known to be polynomially solvable~\cite{FordFulkerson:flow}.
Due to the rich application realm of this problem, many variants and extensions have been investigated.
Some problems ask to partition the graph into more than two parts to disconnect some vertices
such as the \emph{$k$-way cut problem} (the \emph{$k$-cut problem})~\cite{polynominalKCut,KT11},
the \emph{multiterminal cut problem}~\cite{ComplexityMultitermial,X2010} and
 the \emph{multicut problem}~\cite{Multicut:bounded,MR11}.
Some problems are still going to partition the graph into two parts, but with some additional requirements
beyond the disconnectivity.
One of the most extensively studied additional requirements is the constraint
on the numbers of vertices or edges in each of the two parts.
For examples, the \emph{balanced cut problem}~\cite{ARV04,FM06,LR99} and the \emph{minimum bisection problem}~\cite{CLPPS14,FK02,FKN00} require
 the numbers of vertices in the two parts of the cut as close as possible.
The (\emph{balanced}) \emph{judicious bipartition problem}~\cite{Judicious} has conditions on the numbers of edges in the two parts.
Some other well studied additional requirements include conditions on the connectivity of the two parts
 such as the \emph{2-disjoint connected subgraphs problem}~\cite{cppw14},
and conditions on the degree of the two parts, such as the series of bipartition problems
with degree constraints~\cite{BB:partition,decompose:treewidth,decompose:alg,Sdco,XN17}.

In this paper, we study the \emph{bounded-degree cut} problem,
 which belongs to the latter kind of the extensions: to partition a given graph into two parts
with some degree constraints on the induced subgraphs of the two parts.
We mainly consider the upper bounds of the degree.
%For two disjoint vertex subsets $A$ and $B$ in a graph,
%a minimal $(A,B)$-cut is a partition $(V_A,V_B)$ of the vertex set
%such that $A\subseteq V_A$, $B\subseteq V_B$, and for any $A\subseteq V'_A \subsetneqq V_A$
%is called a minimal $(A,B)$-cut
The problem is defined as follows.

\noindent\rule{\linewidth}{0.2mm}
\textsc{bounded-degree cut} (with parameter: $k$)\\
\textbf{Instance:} A multigraph $G=(V,E)$,  two disjoint nonempty vertex subsets $A,B\subseteq V$,
 two  functions  $\mathrm{u}_A$ and $\mathrm{u}_B$ from $V$ to $\{0,1,\ldots,|E|\}$
   and an integer $k\geq 0$.\\
\textbf{Question:}  Does there exist a minimal $(A,B)$-cut $(V_A,V_B)$ such that\\
%$A\subseteq V_A$, $B\subseteq V_B$,\\
the number of edges with one endpoint in $V_A$ and one endpoint in $V_B$ is at most $k$,\\
for each vertex $v\in V_A$, the degree of it in the induced graph $G[V_A]$ is at most $\mathrm{u}_A(v)$, and\\
for  each vertex $v\in V_B$, the degree of it in the induced graph $G[V_B]$ is at most $\mathrm{u}_B(v)$?\\
%it holds that $|E_G(V_A)|\leq k$; \\
%~-~ $\mathrm{deg}_G(v;V_A)\leq  \mathrm{u}_A(v)$ for all vertices $v\in V_A$; and \\
%~-~ $\mathrm{deg}_G(v;V_B)\leq  \mathrm{u}_B(v)$ for all vertices $v\in V_B$?\\
\rule{\linewidth}{0.2mm}
%\medskip

%In fact, many classical NP-hard graph problems can be formulated as a bounded-degree cut problem
% with the objective to optimize the size of one part.
%For examples, the famous \emph{minimum vertex cover problem} is to minimize the size of $V_A$
%of a partition $(V_A,V_B)$ with
%the functions $\mathrm{u}_A(v)=|E|$ and $\mathrm{u}_B(v)=0$ for all vertices $v$;
%The \emph{minimum dominating set problem} is to minimize the size of $V_A$
%of a partition $(V_A,V_B)$ with $\mathrm{u}_A(v)=|E|$ and $\mathrm{u}_B(v)=\mathrm{deg}_G(v)-1$ for all vertices $v$,
%where $\mathrm{deg}_G(v)$ is the degree of $v$ in the original graph $G$;
%The \emph{minimum independent dominating set problem}~\cite{ids} is to minimize the size of $V_A$
%of a partition $(V_A,V_B)$ with $\mathrm{u}_A(v)=0$ and $\mathrm{u}_B(v)=\mathrm{deg}_G(v)-1$
%for all vertices $v$.
% In \textsc{bounded-degree cut}, our target is different.
%It is to minimize the number of edges between the partition, instead of  the size of one part.
%So this problem looks more like a cut problem with additional requirements.

During the last decade, cut related problems were extensively studied
from the viewpoint of parameterized algorithms.
The parameterized complexity of many variants and extensions of the minimum cut problem have be developed.
In this paper, we will study \textsc{bounded-degree cut} from the viewpoint of parameterized algorithms.
The naive brute-force algorithm to enumerate all partitions can solve
 \textsc{bounded-degree cut} in $2^{|V|}\cdot|G|^{O(1)}$ time.
The exponential part of the running time is related to the input size $|V|$.
We show that
this problem admits a \emph{fixed-parameter tractable} (FPT) algorithm with parameter $k$, an
algorithm with running time $f(k)\cdot|G|^{O(1)}$ for a computable function $f(\cdot)$.
Our main result is the first single-exponential FPT algorithm for \textsc{bounded-degree cut}.

\begin{theorem}\label{th:mainresult}
\textsc{bounded-degree cut} admits
an FPT  algorithm that runs in  $2^{18k} \cdot |G|^{O(1)}$ time.
\end{theorem}

This theorem also implies that \textsc{bounded-degree cut} can be solved
in polynomial time for $k=O(\log |G|)$.
%This theorem also implies some other interesting results.
%We look at the \emph{subset independent dominating set} problem.
%Given a graph $G=(V,E)$ and a vertex subset $T\subseteq V$,
%this problem asks  to decide whether
%there is an independent vertex subset $S\subseteq V$ with size $|S|\leq k'$
%such that each vertex in $T$ is either in $S$ or adjacent to a vertex in $S$.
%When $T=V$, this problem becomes the well known independent dominating set problem.
%Since the independent dominating set problem is W[2]-hard with the parameter $k'$~\cite{IDS},
%its generalization, the  subset independent dominating set problem is also W[2]-hard with the parameter $k'$.
%The subset independent dominating set problem can be formulated as a bounded-degree cut problem:
%to find a partition $(V_A,V_B)$ such that $\mathrm{u}_A(v)=0$ for all vertices $v\in V$,
%$\mathrm{u}_B(u')=\mathrm{deg}_G(u')-1$ for all vertices $u'\in T$,
%and $\mathrm{u}_B(u'')=\mathrm{deg}_G(u'')$ for all vertices $u'\in V\setminus T$.
%Thus, Theorem~\ref{th:mainresult} implies that the subset independent dominating set problem is FPT with the parameter $k$ being the number of edges between $S$ and $V\setminus S$.
%More application of Theorem~\ref{th:mainresult} will be shown later.

\subsection{Related work}
There are several interesting contributions on finding a cut or partition of a graph
with some additional requirements. It is known that the minimum ($s,t$)-cut problem is polynomially solvable.
However, the \emph{balanced minimum ($s,t$)-cut problem} is NP-hard~\cite{FM06},
which is to decide whether there is a minimum ($s,t$)-cut such that
 the number of vertices in each part is at most $0<\alpha< 1$ times of the total vertex number.
We can add some trivial vertices in the input graph to make $\alpha$ always being $0.5$.
Note that in this problem, the cut is required to be a minimum ($s,t$)-cut.
Let $k$ denote the size of a minimum ($s,t$)-cut.
By developing a dynamic programming algorithm, Feige and Mahdian~\cite{FM06}
 showed that the vertex-deletion variant of
the balanced minimum ($s,t$)-cut problem is FPT with the parameter $k$.
This algorithm also works for the edge-deletion version.
Another related problem is the (\emph{vertex}) \emph{minimum bisection problem}, which is to find a (vertex) cut of size
at most $k$ such that the two parts of the cut have the same number of vertices.
Marx's result in~\cite{M06} implies that the vertex minimum bisection problem is W[1]-hard with the parameter $k$.
Cygan et.al.~\cite{CLPPS14} showed that the edge vertex version of the minimum bisection problem is FPT.
The above problems have requirements on the vertex numbers in the two parts of the cut or partition.
The \emph{judicious bipartition problem}  requires that the
numbers of edges in the two parts are bounded by $k_1$ and $k_2$ respectively.
Lokshtanov et.al.~\cite{Judicious}
proved that the judicious bipartition problem is FPT with parameter $k_1+k_2$.
In this paper, we consider \textsc{bounded-degree cut}, which is a cut problem with additional requirements on the upper bound of the degree of each vertex in the two parts, and take the cut size as the parameter.

\subsection{Our methods}
The main idea of the algorithm is to construct from a given instance in a graph $G$
a set of at most $2^{18k}$ new ``easy'' instances on the same graph with a special structure
such that (i) the feasibility of each  easy  instances can be tested in $|G|^{O(1)}$ time;
and (ii) the original instance is feasible if and only if
at least one of the  easy instances is feasible.
Constructing such  easy  instances  and
testing the feasibility of all these  give an FPT algorithm for the original instance.
 The idea of converting a general instance  to a set of ``easy'' instances
 has been used to design parameterized algorithms
 for several hard and important problems~\cite{topological,KT11,randomized,CLPPS14}.
 %One of the most important steps is the reduction.
 The construction of easy instances is the most important step.
 Some of the crucial techniques used in our construction is based on
 the concept of \emph{important cut} (or \emph{important separator}) introduced
 by Marx~\cite{M06}.
 Important cuts and separators play an important role in designing FPT algorithms
  for cut problems.
 The fixed-parameter tractability of the (directed) multiterminal cut problem~\cite{CLL2009,CHM2013},
 the multicut problem~\cite{MR11}, the directed feedback vertex set problem~\cite{CLLOR2008,subsetFVS}
 and many other important problems were proved by using important cuts and separators together with some other techniques. We will apply important cuts in a nontrivial way to obtain some general lemmas for bounded sets related
 to cuts. These are crucial for us to design FPT algorithms.

 The framework of our algorithm is as follows.
 For a given instance $(G,A,B,  \mathrm{u}_A, \mathrm{u}_B,k)$
 with a feasible $(A,B)$-cut $(V_1,V_2)$, we try to guess
 some subsets $V'_1\subseteq V_1\setminus A$ and $V'_2\subseteq V_2\setminus B$
 so that the new instance  $(G,A^*=A\cup V'_1, B^*=B\cup V'_2,  \mathrm{u}_A, \mathrm{u}_B,k)$
 remains feasible and is an ``easy'' instance in the sense that the feasibility can be tested in $n^{O(1)}$ time.
 We call a vertex $v$ in $G$
   \emph{$A$-unsatisfied} (resp., \emph{$B$-unsatisfied})
 if its degree in   $G$ is greater than
 $\mathrm{u}_A(v)$ (resp., $\mathrm{u}_B(v)$), and call an $A$- or $B$-unsatisfied vertex
 \emph{unsatisfied}.
 We first guess whether each unsatisfied vertex belongs to $V_1$ or $V_2$.
 % for a feasible $(A,B)$-cut $(V_1,V_2)$.
 Although the number of  unsatisfied vertices may not be bounded by a function of $k$,
 the set $Z_{A1}$ of $A$-unsatisfied vertices in $V_1$ can contain at most $k$ vertices,
 because each vertex in $Z_{A1}$ must be adjacent to a vertex in $V_2$ to satisfy
 the degree constraint.
 Symmetrically
 the set $Z_{B2}$ of $B$-unsatisfied vertices in $V_2$ can contain at most $k$ vertices.
 By applying the result on important cuts and our new lemmas, we construct
 at most $2^{12k}$ pairs $(X_1,X_2)$ of vertex subsets one of which is equal to $(Z_{A1},Z_{B2})$.
 For the set $(X_1,X_2)=(Z_{A1},Z_{B2})$ and the set $Z_{A2}$ (resp., $Z_{B1}$)
 of $A$-unsatisfied vertices in $V_2$ (resp.,  $B$-unsatisfied vertices in $V_1$),
 we see that the new instance
  $(G, A'=A\cup Z_{A1}\cup Z_{B1},  B'=B\cup Z_{A1}\cup Z_{B2},  \mathrm{u}_A, \mathrm{u}_B,k)$
  remains feasible.
  However, this instance may not be ``easy'' yet in our sense, because
  whether  the degree constraint on a vertex in $Z_{A1}$ or $Z_{B2}$ holds or not
    depends  on a choice of an $(A',B')$-cut in the new instance.
 We next guess whether each neighbor of a vertex in  $Z_{A1}\cup Z_{B2}$
 belongs to $V_1$ or $V_2$.
 %  for a feasible $(A,B)$-cut $(V_1,V_2)$.
 We see that the set $W_{B1}$ of neighbors of $Z_{B2}$ belonging to  $V_1$ can contain at most $k$ vertices,
 because the number of such neighbors is bounded from above by the cut-size of $(V_1,V_2)$.
 Symmetrically
 the set $W_{A2}$ of neighbors of $Z_{A1}$ belonging to  $V_2$   contains at most $k$ vertices.
 By applying the result on important cuts and our new lemmas again, we construct
 at most $2^{6k}$ pairs $(Y_1,Y_2)$ of vertex subsets one of which is equal to $(W_{B1},W_{A2})$.
 Let $W_{B2}$ (resp., $W_{A1}$) denote of neighbors of $Z_{B2}$ belonging to  $V_2$
 (resp., of $Z_{A1}$ belonging to  $V_1$).
 Then for the right choice $(X_1,X_2)=(Z_{A1},Z_{B2})$ and $(Y_1,Y_2)=(W_{B1},W_{A2})$,
 the resulting instance $(G, A^*= A\cup Z_{A1}\cup Z_{B1}\cup W_{A1}\cup W_{B1},
 B^*=B\cup  Z_{A1}\cup Z_{B2}\cup W_{A2}\cup W_{B2},    \mathrm{u}_A, \mathrm{u}_B,k)$
 remains feasible and is an easy instance where  the degree constraint on a vertex in $Z_{A1}$ or $Z_{B2}$ holds or not
  does not  depend   on a choice of an $(A^*,B^*)$-cut in the new instance.

 The remaining part of the paper is organized as follows.
Section~\ref{sec:prelimi} reviews basic notations on graphs and cuts
and properties on minimum cuts and important cuts.
Section~\ref{sec:candidate}   introduces some technical lemmas, which will be   building blocks
of our algorithm. We believe that these lemmas can be used to design FPT algorithms for more problems.
Section~\ref{sec:easy_case} defines ``easy'' instances and proves the polynomial-time
solvability.
Based on the results in Section~\ref{sec:candidate},
Section~\ref{sec:branching} describes how to generate candidate set pairs
$(X_1,X_2)$ for the pair $(Z_{A1},Z_{B2})$ and $(Y_1,Y_2)$ for the pair $(W_{B1},W_{A2})$.
%Section~\ref{sec:hyperedge} addresses an application of
%\textsc{bounded-degree cut} to a problem of removing hyperedges
%from a hypergraph so as to satisfy a degree upper bound condition.
Finally Section~\ref{sec:conclusion} makes some concluding remarks.

\section{Preliminaries}\label{sec:prelimi}

In this paper, a graph $G=(V,E)$ stands for an undirected multigraph
with a vertex set $V$ and an edge set $E$.
We will use $n$ and $m$ to denote the sizes of $V$ and $E$, respectively.
%Let $G=(V,E)$ be a graph,   $X$ be a subset of $V$, and $S,T$ be two disjoint subsets of $V$.
Let $X$ be a subset of $V$.
We use $G-X$ to denote
the graph obtained from $G$ by removing vertices in $X$
together with all edges incident to vertices in $X$.
% , where $G-\{v\}$ for a vertex $v$ may be written as $G-v$.
Let $G[X]$ denote the graph induced by  $X$,
i.e., $G[X]= G -(V\setminus X)$.
Let $N_G(v)$ denote the set  of neighbors of a vertex $v$ in $G$,
and let $N_G(v;X)\triangleq N_G(v)\cap X$.
Let $N_G(X)$ denote the set of neighbors $u\in V\setminus X$ of a vertex $v\in X$,
i.e., $N_G(X)= \bigcup_{v\in X}N_G(v;V\setminus X)$.
%Let $N_G[X]\triangleq N_G(X)\cup X$.
For two disjoint vertex subsets $X$ and $Y$,
the set of edges with one endpoint in $X$ and one endpoint in $X$ is denoted by $E_G(X,Y)$,
and $E_G(X,V\setminus X)$ may be simply written as $E_G(X)$.
Define $\mathrm{deg}_G(v)\triangleq |E_G(\{v\})|$ and
 $\mathrm{deg}_G(v;X)\triangleq |E_{G}(\{v\},X\setminus \{v\})|$.

\begin{definition}
\emph{\textbf{\emph{($(S,T)$-cuts)}}
For two disjoint vertex subsets $S$ and $T$,
an ordered pair $(V_1,V_2=V\setminus V_1)$ is called an {\em $(S,T)$-cut} if $S\subseteq V_1$ and $T\subseteq V_2$,
and its  {\em cut-size}  is defined to be $|E_G(V_1)|$.}
\end{definition}

\begin{definition} \emph{\textbf{\emph{(minimal $(S,T)$-cuts, minimum $(S,T)$-cuts and MM $(S,T)$-cuts)}}
An $(S,T)$-cut $(V_1,V_2)$ is minimal if
$E_G(V_1)$ does not contain $E_G(V'_1)$ or $E_G(V'_2)$ as a subset for any $S\subseteq V'_1 \subsetneqq V_1$ and $T\subseteq V'_2 \subsetneqq V_2$.
An $(S,T)$-cut $(V_1,V_2)$ is minimum if
its cut-size $|E_G(V_1,V_2)|$ is minimum  over all $(S,T)$-cuts.
An $(S,T)$-cut $(V_1,V_2)$ is called an MM $(S,T)$-cut if
it is a minimum $(S,T)$-cut such that  $|V_1|$ is maximum  over all minimum $(S,T)$-cuts.}
\end{definition}

\begin{lemma}\label{lm:min_cut} \emph{(\cite{FordFulkerson:flow,ComplexityMultitermial})}
For   two disjoint vertex subsets $S, T\subseteq V$, an MM $(S,T)$-cut is unique
and it can be found in $O(\min\{n^{2/3},m^{1/2}\}m)$ time.
\end{lemma}

\begin{definition}\emph{ \textbf{\emph{(important cuts)}}
%For an integer $k\geq 0$,
A minimal $(S,T)$-cut $(X,V\setminus X)$ is called
  an important $(S,T)$-cut if
%$|E_G(X)|\leq k$ and
there is no   $(S,T)$-cut $(X',V\setminus X')$  such that $X'\supsetneqq X$ and
$|E_G(X')| \leq |E_G(X)|$.}
\end{definition}

The following result is known \cite{CLL2009,MR11}.

\begin{lemma}   \label{lem_4kimp}
Let $S, T\subseteq V$ be non-empty subsets in a graph $G=(V,E)$.\\
{\rm (i)} For any subset $X$ with $S\subseteq X\subseteq V\setminus T$,
the MM $(X,T)$-cut is an important $(S,T)$-cut;\\
{\rm (ii)}
There are at most $4^k$ important $(S,T)$-cuts of size at most $k$ and one
can list all of them in $4^k (n+m)^{O(1)}$ time.
% $O(4^k k(n+m))$ time. [our graph is multigraphs]
\end{lemma}

\section{Candidate Sets}\label{sec:candidate}%%%%%%%%%%%%%%%%%
We introduce the next technical lemmas, which will be used to  build blocks
of our algorithm. These lemmas are crucial for us to design FPT algorithms.

\begin{lemma}   \label{lem_candidate1}
Let  $A,B,C\subseteq V$   be non-empty subsets in a graph $G=(V,E)$  and $k$ and
 % $\ell~(< n)$
 $\ell$ be nonnegative integers.
Then one can find  in $2^{3(k+\ell)}(n+m)^{O(1)}$ time
a family $\mathcal{X}$ of at most  $2^{3(k+\ell)}$  subsets of $C$
 with a property that $C\cap V_1\in \mathcal{X}$ for any minimal $(A, B)$-cut $(V_1,V_2)$
  with size at most $k$ such that
 $| C\cap V_1|\leq \ell$.
\end{lemma}
\pf{ Let $\mathrm{Cut}(A,B,C,k,\ell;G)$ denote the set of minimal
$(A, B)$-cuts $(V_1,V_2)$ in $G$  with size at most $k$ such that  $| C\cap V_1|\leq \ell$.
 Construct a multigraph $H_b$   from $G$ by choosing a vertex $b\in B$
and adding a new edge
 between $b$  and each vertex $u\in C$,
 and let   $\mathrm{ICut}(A,B,k+\ell;H_b)$  denote  the set of
  important $(A,B)$-cuts  in $H_b$ of size at most   $k+\ell$.
 By Lemma~\ref{lem_4kimp}(ii),
    $|\mathrm{ICut}(A,B,k+\ell;H_b)|\leq  4^{k+\ell}$  holds,
    and   $\mathrm{ICut}(A,B,k+\ell;H_b)$ can be found in time
  $4^{k+\ell} (n+m)^{O(1)}$.
  %= 4^{k+\ell}   n^{O(1)}$ time,   where  $\ell<n$.

 For any minimal $(A,B)$-cut $(V_1,V_2)\in  \mathrm{Cut}(A,B,C,k,\ell;G)$,
 we see by Lemma~\ref{lem_4kimp}(i)
 that
 the MM $(A\cup (C\cap V_1), B)$-cut $(S,T)$
 in $H_b$ is an important $(A,B)$-cut  in $H_b$ of size at most
  $k+| C\cap V_1|\leq k+\ell$, where
$ C\cap V_1 \subseteq N_{H_b}(b)\cap S$ holds.
Construct the family $\mathcal{X}$ of subsets
\[\mbox{  $X\subseteq  N_{H_b}(b)\cap S$ with $|X|\leq \ell$
for each $(A,B)$-cut $(S,T)\in  \mathrm{ICut}(A,B,k+\ell;H_b)$. }\]
Then $\mathcal{X}$
contains the set $C\cap V_1$ for each  $(A, B)$-cut $(V_1, V_2)\in \mathrm{Cut}(A,B,C,k,\ell;G)$.

For each  important $(A,B)$-cut $(S,T)\in \mathrm{ICut}(A,B,k+\ell;H_b)$,
the family $\mathcal{X}$ contains at most
$$\sum_{i=0}^\ell{|N_{H_b}(b)\cap S|\choose i}
 \leq \sum_{i=0}^\ell{k+\ell \choose i}<2^{k+\ell}$$
 subsets $X$.
 Since $|\mathrm{ICut}(A,B,k+\ell;H_b)|\leq 4^{k+\ell}$,
 it holds that $|\mathcal{X}|\leq 4^{k+\ell}\cdot 2^{k+\ell}=2^{3(k+\ell)}$
 and the family $\mathcal{X}$ can be constructed in
 $4^{k+\ell}  (n+m)^{O(1)}+  2^{3(k+\ell)}(n+m)^{O(1)}$ time.
 This proves the lemma.
}\bigskip

% Such a family $\mathcal{X}$ is called a {\em candidate family} of $(A,B,C,k,\ell)$
% and a set in  $\mathcal{X}$ is called a {\em candidate} to the set $C\cap V_1$.

\begin{lemma}   \label{lem_candidate2}
Let  $A,B,B'\subseteq V$ be non-empty subsets in a graph $G=(V,E)$, where $B'\subseteq B$,
 and $k$ be a nonnegative integer.
 Then one can find  in $2^{3k}(n+m)^{O(1)}$ time
 a family $\mathcal{Y}$ of at most  $2^{3k}$  subsets of $N_G(B')$
 with a property  that $N_G(B')\cap V_1\in \mathcal{Y}$
 for any  minimal $(A, B)$-cut
  $(V_1,V_2)$  with size at most $k$.
\end{lemma}
\pf{Let $\mathrm{Cut}(A,B,k)$ denote the set of minimal
$(A, B)$-cuts in $G$  with size at most $k$.
 Let   $\mathrm{ICut}(A,B,k)$ denote  the set of
  important $(A,B)$-cuts  in $G$ with size at most   $k$.
 By Lemma~\ref{lem_4kimp}(ii),
    $|\mathrm{ICut}(A,B,k)|\leq  4^{k}$  holds,
    and   $\mathrm{ICut}(A,B,k)$ can be found in
  $4^{k} n^{O(1)}$ time.

 For any minimal $(A,B)$-cut $(V_1,V_2)\in  \mathrm{Cut}(A,B,k)$,
 we see by Lemma~\ref{lem_4kimp}(i)
 that
 the MM $(A  \cup (N_G(B')\cap V_1), B)$-cut $(S,T)$
   is an important $(A,B)$-cut  of size at most
  $k$, where
$ N_G(B')\cap V_1 \subseteq N_G(B')\cap S$ holds.
Construct  the family $\mathcal{Y}$ of subsets
\[\mbox{  $Y\subseteq N_G(B')\cap S$
for each $(A,B)$-cut $(S,T)\in  \mathrm{ICut}(A,B,k)$.}\]
Then  $\mathcal{Y}$ contains the set $N_G(B')\cap V_1$
for each  $(A, B)$-cut $(V_1, V_2)\in \mathrm{Cut}(A,B,k)$.

Note that the size of $N_G(B')\cap S$ is at most the size of the cut $(S,T)$.
For each  important $(A,B)$-cut $(S,T)\in \mathrm{ICut}(A,B,k)$,
the family $\mathcal{Y}$ contains at most
$$2^{|N_G(B')\cap S|}\leq 2^k$$
 subsets $Y$.
 Since $|\mathrm{ICut}(A,B,k )|\leq 4^{k}$,
 it holds that $|\mathcal{Y}|\leq 4^{k }\cdot 2^{k }=2^{3k}$
 and the family $\mathcal{Y}$ can be constructed in
 $4^{k }  (n+m)^{O(1)}+  2^{3k}(n+m)^{O(1)}$ time.
 This proves the lemma.
}\bigskip

%Such a family $\mathcal{Y}$ is called a {\em candidate family} of $(A,B,D,k,\ell)$
%and a set in  $\mathcal{Y}$ is called a {\em candidate} to the set $N_G(D)\cap V_2$.

\section{Restriction to an Easy Case}\label{sec:easy_case}%%%%%%%%%%%%%%%

 Recall that an instance of \textsc{bounded-degree cut}
is defined by a tuple
$(G=(V,E),A,B , \mathrm{u}_A, \mathrm{u}_B, k)$
such that  $G$ is a multigraph, $A, B\subseteq V$ are two disjoint vertex subsets,
$\mathrm{u}_A$ and $\mathrm{u}_B$ are two functions from $V$ to $\{0,1,\ldots,|E|\}$, and $k$ is a nonnegative integer.
We will use $I=(G=(V,E),A,B)$ to denote an instance of the problem,
where $\mathrm{u}_A$, $\mathrm{u}_B$ and $k$ are omitted since
they remain unchanged throughout our argument.
We call a minimal $(A,B)$-cut $(V_A,V_B=V\setminus V_A)$   {\em feasible} to an instance $I$
if   \\
~-~ $|E_G(V_A)|\leq k$; \\
~-~ $\mathrm{deg}_G(v;V_A)\leq  \mathrm{u}_A(v)$ for all vertices $v\in V_A$; and \\
~-~ $\mathrm{deg}_G(v;V_B)\leq  \mathrm{u}_B(v)$ for all vertices $v\in V_B$,\\
where the last two conditions are  also called the \emph{degree constraint}.
A feasible $(A,B)$-cut in $I$ is called a {\em solution} to $I$,
and an instance $I$ is called {\em feasible} if it admits a solution.
\medskip

\textsc{bounded-degree cut} is to decide whether a given instance $I$ is
 feasible  or not.
We observe the next.

\begin{lemma} \label{lem_restriction}
For an instance $I=(G=(V,E),A,B)$
and disjoint nonempty subsets $X,Y\subseteq V$, let $I_{X,Y}$ denote
the instance   $(G, A\cup X ,B\cup Y)$.  \\
{\rm (i)}
If $I$ is infeasible, then $I_{X,Y}$ is infeasible for any $X,Y\subseteq V$; \\
{\rm (ii)}
If $I$ is  feasible and
$X\subseteq V_A$ and $Y\subseteq V_B$ hold for a feasible $(A,B)$-cut $(V_A,V_B)$ to $I$,
then $I_{X,Y}$ admits a feasible $(A\cup X, B\cup Y)$-cut, which is also feasible to $I$.
\end{lemma}
\pf{{\rm (i)} Assume to the contrary that $I_{X,Y}$ is feasible and
$(V_A,V_B)$ is a feasible $(A\cup X ,B\cup Y)$-cut. Then it holds that $A\subseteq V_A$ and $B\subseteq V_B$
and the cut $(V_A,V_B)$ satisfies the conditions in the definition of feasible cuts. Thus,
$(V_A,V_B)$ is also a feasible $(A,B)$-cut, a contradiction to the fact that  $I$ is infeasible.

{\rm (ii)} First of all, it is clear that $(V_A,V_B)$ is still a feasible $(A\cup X, B\cup Y)$-cut.
Then $I_{X,Y}$ admits feasible $(A\cup X, B\cup Y)$-cuts. Let $(V'_A,V'_B)$ be an arbitrary feasible $(A\cup X, B\cup Y)$-cut. The above proof for {\rm (i)} shows that $(V'_A,V'_B)$ is also a feasible $(A, B)$-cut.
}

\medskip

Let $I=(G,A,B)$ be an instance.
We use $Z_A$ and $Z_B$ to denote the sets of $A$-unsatisfied vertices and $B$-unsatisfied vertices, respectively,  i.e.,
\[\mbox{
$Z_A\triangleq \{v\in V\mid   \mathrm{deg}_G(v)>  \mathrm{u}_A(v)\}$ and
$Z_B\triangleq \{v\in V\mid   \mathrm{deg}_G(v)>  \mathrm{u}_B(v)\}$, }\]
where  $Z_A\cap Z_B$ may not be empty.
%Recall that a vertex $v$ in the graph is called \emph{$A$-unsatisfied} (resp., \emph{$B$-unsatisfied}) if its degree in the graph $G$ is greater than $\mathrm{u}_A(v)$ (resp., $\mathrm{u}_B(v)$).
% Recall that $U=V\setminus (A\cup B)$.
We call $I$  an \emph{easy instance} if it holds that \\
1. $Z_A\cup Z_B \subseteq A\cup B$, \\
2. $N_G(Z_A\cap A) \subseteq A\cup B$, and\\
3. $N_G(Z_B\cap B) \subseteq A\cup B$.

\begin{lemma} \label{lem_easy}
The feasibility of an easy instance of
\textsc{bounded-degree cut}  can be tested in $(n+m)^{O(1)}$ time.
\end{lemma}
\pf{Let $I= (G,A,B)$ be an easy instance.
Note that the degree constraint to each vertex in $V\setminus (A\cup B)
\subseteq V\setminus (Z_A\cup Z_B)$ is satisfied for any $(A,B)$-cut in $I$.
First we test in $n^{O(1)}$ time
whether there is a vertex $v\in Z_A\cap A$ with
$\mathrm{deg}_G(v;A)> \mathrm{u}_A(v)$
(resp.,    $v\in Z_B\cap B$ with
$\mathrm{deg}_G(v;B)> \mathrm{u}_B(v)$) or not.
If so,
then clearly $I$ admits no solution.
Assume that no such vertices exist in $I$.
Then each $v\in Z_A$ satisfies $v\not \in A$
or $v\in Z_A\cap A$.
In the former, it holds that $v\in B$ since $Z_A\subseteq A\cup B$,
where we do not need to consider the degree constraint by  $\mathrm{u}_A(v)$.
In the latter,  it holds that $N_G(v)\subseteq A\cup B$
 since $N_G(Z_A\cap A) \subseteq A\cup B$,
where $\mathrm{deg}_G(v;V_1)=\mathrm{deg}_G(v;A)\leq \mathrm{u}_A(v)$
for any $(A,B)$-cut $(V_1,V_2)$ in $I$,
satisfying the degree constraint by  $\mathrm{u}_A(v)$.
Analogously no vertex in $Z_B$ violates
the degree constraint by  $\mathrm{u}_B(v)$ for any choice of
  $(A,B)$-cuts $(V_1,V_2)$ in $I$.

Now $I$ admits a solution if and only if it has an $(A,B)$-cut with size at most $k$,
which can be checked in $(n+m)^{O(1)}$ time by Lemma~\ref{lm:min_cut}.
This proves the lemma.
}\bigskip

We will construct from a given instance at most $2^{18k}$ easy instances so that
where the original instance is feasible  if and only if
at least one of the easy instances is feasible.

\section{Constructing Easy Instances}\label{sec:branching}%%%%%%%%%
\begin{figure}[htbp] \begin{center}
\includegraphics[scale=0.25]{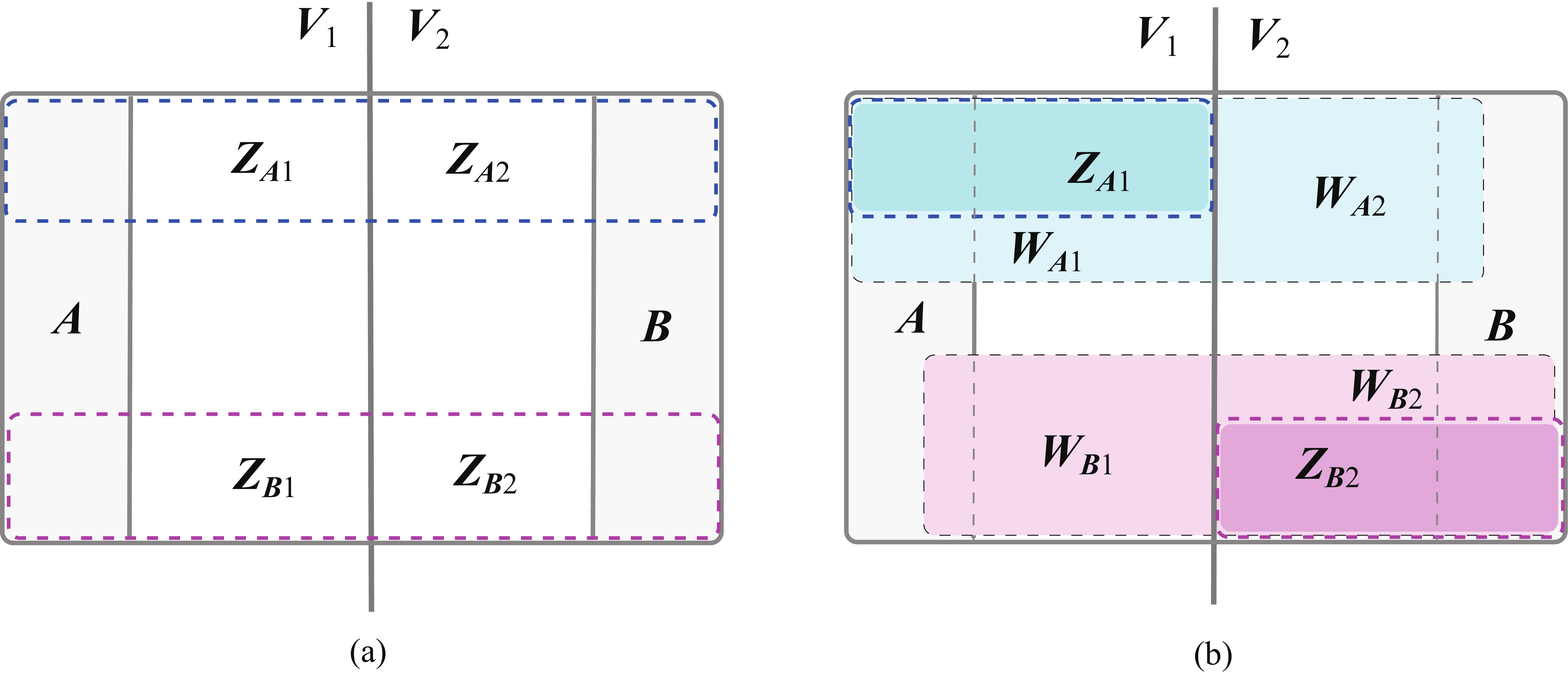} \end{center}
\caption{(a) An $(A,B)$-cut $\pi=(V_A,V_B)$ to $I$
and the partitions $\{ Z_{A1},Z_{A2}\}$ of $Z_A$
and $\{ Z_{B1},Z_{B2}\}$ of $Z_B$ by $\pi$,
where possibly $Z_A\cap Z_B\neq\emptyset$,
(b) The partitions $\{W_{B1},W_{B2}\}$ of $W_B=N_G(Z_{B2})$
and $\{W_{A1},W_{A2}\}$ of $W_A=N_G(Z_{A1})$ by $\pi$,
where possibly $W_A\cap W_B\neq\emptyset$.
}
\label{fi:degree-cut}
\end{figure}

For a minimal   $(A,B)$-cut $\pi=(V_1,V_2)$ (not necessarily feasible) in
a given  instance $I=(G=(V,E),A,B)$,
we define the following notation on vertex subsets: \\
~~~ $Z_{Ai}\triangleq Z_A \cap V_i$ and $Z_{Bi}\triangleq Z_B \cap V_i$, $i=1,2$; \\
~~~  $W_A\triangleq N_G(Z_{A1})$  and  $W_B\triangleq N_G(Z_{B2})$;
 $W_{Ai}\triangleq W_A \cap V_i$, and $W_{Bi}\triangleq W_B \cap V_i$ , $i=1,2$;\\
~~~  $A_{\pi}\triangleq A\cup Z_{A1} \cup Z_{B1} \cup W_{A1} \cup W_{B1}$ and
$B_{\pi}\triangleq B\cup Z_{A2} \cup Z_{B2} \cup W_{A2} \cup W_{B2}$. \\
See  in Fig.~\ref{fi:degree-cut} for an illustration on these subsets.
Observe that the resulting instance $(G,A_{\pi},B_{\pi})$ is an easy instance.
By Lemma~\ref{lem_restriction},
the $(A,B)$-cut $\pi=(V_1,V_2)$ is feasible if and only if
 the corresponding  instance $(G,A_{\pi},B_{\pi})$ is feasible.

\subsection{Partitioning Unsatisfied Vertices}

For a minimal   $(A,B)$-cut $(V_1,V_2)$ to an instance $I$,
let  $Z_{A1}$ and $Z_{B2}$ be the subsets defined in the above.
We observe that if the cut is feasible, then
 $$|Z_{A1}|,|Z_{B2}| \leq k$$ since each vertex in $Z_{A1}\cup  Z_{B2}$
 has at least one incident edge included in $E_G(V_1,V_2)$
 so that the degree constraint on the vertex holds.

By applying Lemma~\ref{lem_candidate1} to $(A,B,C=Z_A,k,\ell=k)$,
we can construct  in $2^{6k}   (n+m)^{O(1)}$ time
a family $\mathcal{X}_1$ of at most $2^{6k}$ subsets of $Z_A$
such that
$\mathcal{X}_1$  contains the set
$Z_{A1} $  defined to each feasible $(A,B)$-cut
$(V_1,V_2)$ in the instance $I=(G,A,B)$.
Symmetrically it takes
 $2^{6k}   (n+m)^{O(1)}$ time to find a family
 $\mathcal{X}_2$ of
at most $2^{6k}$ subsets of  $Z_B$ such that   $\mathcal{X}_2$
contains the set   $Z_{B2}$  defined to each feasible $(A,B)$-cut
$(V_1,V_2)$ in the instance $I=(G,A,B)$.
Then the set $\mathcal{X}_{1,2}$ of all pairs $(X_1,X_2)$ of disjoint sets $X_i\in \mathcal{X}_i$, $i=1,2$
contains the pair $(Z_{A1},Z_{B2})$
defined to each feasible $(A,B)$-cut
$(V_1,V_2)$ in   $I$.
By noting that $|\mathcal{X}_{1,2}|\leq 2^{6k}2^{6k}= 2^{12k}$,
we obtain the next.

\begin{lemma} \label{lem_step1}
 Given an instance $I=(G, A,B)$,
one can construct in $ 2^{12k}(n+m)^{O(1)}$ time
at most $2^{12k}$ new instances $I'=(G, A',B')$ with $Z_A\cup Z_B \subseteq A' \cup B'$,
one of which is equal to  $(G, A\cup Z_{A1}\cup Z_{B1}, B \cup Z_{A2} \cup Z_{B2})$
for each  feasible $(A,B)$-cut $(V_1,V_2)$ to $I$.
\end{lemma}
%\pf{The proof to be added here.}

\subsection{Partitioning Neighbors of Unsatisfied Vertices}

For a minimal $(A,B)$-cut $(V_1,V_2)$ to an instance $I$,
let  $W_{A2}$ and $W_{B1}$ be the subsets defined in the above.
We observe that if the cut is feasible, then  $$   |W_{B1}| , |W_{A2}| \leq k $$
since each of   $|N_G(Z_{B2})\cap V_1|$ and
$|N_G(Z_{A1})\cap V_2|$  is at most $|E_G(V_1,V_2)|\leq k$
to the feasible $(A,B)$-cut $(V_1,V_2)$.

By applying Lemma~\ref{lem_candidate2} to
$(A\cup Z_{A1}\cup Z_{B1},  B\cup Z_{A2}\cup Z_{B2}, B'=Z_{B2}, k)$,
% XXX: D=N_G(Z_{B2})  -->  B'= Z_{B2}
we can construct  in $2^{3k}   n^{O(1)}$ time
a family $\mathcal{Y}_1$ of at most $2^{3k}$ subsets of $N_G(Z_{B2})$
such that
$\mathcal{Y}_1$  contains the set
$W_{B1}=N_G(Z_{B2})\cap V_1$  defined to each feasible $(A,B)$-cut
$(V_1,V_2)$ in the instance $I=(G,A,B)$.
Symmetrically it takes    $2^{3k}   (n+m)^{O(1)}$ time to find
a family $\mathcal{Y}_2$ of at most $2^{3k}$ subsets of $N_G(Z_{A1})$
such that
$\mathcal{Y}_2$  contains the set
$W_{A2}=N_G(Z_{A1})\cap V_2$  defined to each feasible $(A,B)$-cut
$(V_1,V_2)$ in   $I$.
Then the set $\mathcal{Y}_{1,2}$ of all pairs $(Y_1,Y_2)$ of disjoint sets
  $Y_i\in \mathcal{Y}_i$, $i=1,2$
% XXX: $Y_i\in \mathcal{X}_i$,   -->   $Y_i\in \mathcal{Y}_i$,
contains the pair $(W_{B1},W_{A2})$
defined to each feasible $(A,B)$-cut
$(V_1,V_2)$ in the instance $I=(G,A,B)$.
By noting that $|\mathcal{Y}_{1,2}|\leq 2^{6k}$, we obtain the next.

\begin{lemma} \label{lem_step2}
 Given an instance $I=(G, A,B)$ and the subsets $Z_{A1}$ and $Z_{B2}$
 defined to a feasible $(A,B)$-cut
$(V_1,V_2)$ in $I$,
one can construct in $ 2^{6k} (n+m)^{O(1)}$ time
at most $2^{6k}$ new easy
instances $I'=(G, A',B')$,
one of which is equal to  $(G, A_{\pi}, B_{\pi})$ defined to the feasible $(A,B)$-cut
$\pi=(V_1,V_2)$.
\end{lemma}

By Lemmas~\ref{lem_step1} and \ref{lem_step2},
we obtain the next.

\begin{lemma} \label{lem_all_steps}
 Given an instance $I=(G, A,B)$,
one can construct in $ 2^{18k} (n+m)^{O(1)}$ time
at most $2^{18k}$ new easy
instances $I'=(G, A',B')$,
one of which is equal to  $(G, A_{\pi}, B_{\pi})$ for each  feasible $(A,B)$-cut $\pi=(V_1,V_2)$ to $I$.
\end{lemma}

This and Lemma~\ref{lem_easy} imply Theorem~\ref{th:mainresult}.

\section{Concluding Remarks}\label{sec:conclusion}
Cut and partition problems are important problems that have been extensively studied from the viewpoint of
FPT algorithms. In this paper, we study a cut problem with additional constraints on the vertex degree of the two parts of the cut and design the first FPT algorithm for this problem.
To obtain the FPT algorithm, we develop two new lemmas that are based on important cuts.
Important cuts show some properties of bounded-size cuts, while the new lemmas further reveal
some properties of vertex subsets of one part of a bounded-size cut.
We believe these lemmas can be used to design FPT algorithms for more problems.

In \textsc{bounded-degree cut}, we are going to check whether there is a minimal  $(A,B)$-cut
satisfying both the degree constraint and size constraint of most $k$.
We also consider the \textsc{bounded-degree bipartition} problem, which is to check whether there is
$(A,B)$-cut of size at most $k$ satisfying the degree constraint, without the requirement of being minimal.
Note that some $(A,B)$-cuts of size at most $k$ satisfying the degree constraint may not be minimal.
This kind of cuts are not solutions to \textsc{bounded-degree cut}, but are solutions to \textsc{bounded-degree bipartition}. To solve \textsc{bounded-degree bipartition}, we need some techniques more, which will be introduced
in our further work.

%%
%% Bibliography
%%

%% Please use bibtex,

\bibliography{my}

\end{document}